\documentclass[twocolumn,amsmath,amssymb,prl,color,superscriptaddress]{revtex4}
\usepackage{graphicx}
\usepackage{color}

\usepackage{graphicx}
\usepackage{dcolumn}
\usepackage{bm}
\usepackage[T1]{fontenc}
\usepackage{hyperref}
\hypersetup{
     colorlinks = true,
     citecolor  = blue,  
     linkcolor  = magenta 
}
\begin{document}

\title{Monoclinic EuSn$_2$As$_2$: A Novel High-Pressure Network Structure }

\author{Lin Zhao}
\thanks{These authors contribute equally to this work.}
\affiliation{Beijing National Laboratory for Condensed Matter Physics and Institute of Physics, Chinese Academy of Sciences, Beijing 100190, China}
\affiliation{State Key Laboratory of Superhard Materials, Jilin University, Changchun 130012, China}

\author{Changjiang Yi}
\thanks{These authors contribute equally to this work.}
\affiliation{Beijing National Laboratory for Condensed Matter Physics and Institute of Physics, Chinese Academy of Sciences, Beijing 100190, China}

\author{Chang-Tian Wang}
\thanks{These authors contribute equally to this work.}
\affiliation{Beijing National Laboratory for Condensed Matter Physics and Institute of Physics, Chinese Academy of Sciences, Beijing 100190, China}
\affiliation{School of Physical Sciences, University of Chinese Academy of Sciences, Beijing 100049, China}

\author{Zhenhua Chi}
\thanks{These authors contribute equally to this work.}
\affiliation{Key Laboratory of Materials Physics, Institute of Solid State Physics, HFIPS, Chinese Academy of Sciences, Hefei 230031, China}

\author{Yunyu Yin}
\affiliation{Beijing National Laboratory for Condensed Matter Physics and Institute of Physics, Chinese Academy of Sciences, Beijing 100190, China}

\author{Xiaoli Ma}
\affiliation{Beijing National Laboratory for Condensed Matter Physics and Institute of Physics, Chinese Academy of Sciences, Beijing 100190, China}

\author{Jianhong Dai}
\affiliation{Beijing National Laboratory for Condensed Matter Physics and Institute of Physics, Chinese Academy of Sciences, Beijing 100190, China}

\author{Pengtao Yang}
\affiliation{Beijing National Laboratory for Condensed Matter Physics and Institute of Physics, Chinese Academy of Sciences, Beijing 100190, China}

\author{Binbin Yue}
\affiliation{Center for High Pressure Science $\&$ Technology Advanced Research, Haidian, Beijing 100094, China}

\author{Jinguang Cheng}
\affiliation{Beijing National Laboratory for Condensed Matter Physics and Institute of Physics, Chinese Academy of Sciences, Beijing 100190, China}

\author{Fang Hong}
\affiliation{Beijing National Laboratory for Condensed Matter Physics and Institute of Physics, Chinese Academy of Sciences, Beijing 100190, China}

\author{Jian-Tao Wang}
\email{wjt@aphy.iphy.ac.cn}
\affiliation{Beijing National Laboratory for Condensed Matter Physics and Institute of Physics, Chinese Academy of Sciences, Beijing 100190, China}
\affiliation{School of Physical Sciences, University of Chinese Academy of Sciences, Beijing 100049, China}
\affiliation{Songshan Lake Materials Laboratory, Dongguan, Guangdong 523808, China}

\author{Yonghao Han}
\email{hanyh@jlu.edu.cn}
\affiliation{State Key Laboratory of Superhard Materials, Jilin University, Changchun 130012, China}

\author{Youguo Shi}
\email{ygshi@iphy.ac.cn}
\affiliation{Beijing National Laboratory for Condensed Matter Physics and Institute of Physics, Chinese Academy of Sciences, Beijing 100190, China}
\affiliation{Center of Materials Science and Optoelectronics Engineering, University of Chinese Academy of Sciences, Beijing 100049, China}

\author{Xiaohui Yu}
\email{yuxh@iphy.ac.cn}
\affiliation{Beijing National Laboratory for Condensed Matter Physics and Institute of Physics, Chinese Academy of Sciences, Beijing 100190, China}

\date{\today}

\begin{abstract}
The layered crystal of EuSn$_2$As$_2$ has a Bi$_2$Te$_3$-type structure in rhombohedral ($R\bar{3}m$) symmetry and has been confirmed to be an intrinsic magnetic topological insulator at ambient conditions.
Combining {\it ab initio} calculations and \emph{in-situ} x-ray diffraction measurements, we identify a new monoclinic EuSn$_2$As$_2$ structure in $C2/m$ symmetry above $\sim$14 GPa.
It has a three-dimensional network made up of honeycomb-like Sn sheets and zigzag As chains,
transformed from the layered EuSn$_2$As$_2$ via a two-stage reconstruction mechanism
with the connecting of Sn-Sn and As-As atoms successively between the buckled SnAs layers.
Its dynamic structural stability has been verified by phonon mode analysis.
Electrical resistance measurements reveal an insulator-metal-superconductor transition at low temperature around 5 and 15 GPa, respectively, according to the structural conversion, and the superconductivity with a \textit{T}${\rm {_C}}$ value of $\sim 4$ K is observed up to 30.8 GPa.
These results establish a high-pressure EuSn$_2$As$_2$ phase with intriguing structural and electronic properties and expand our understandings about the layered magnetic topological insulators.

\end{abstract}
\maketitle

Topological insulators have attracted much research interest due to their novel band structure and quantum phenomenon \cite{1Hasan,2QiXiaoLiang,3FuLiang,4Moore,5ZhangHaijun,6FuLiang,7Kane}. In the past few years, a good deal of attention has been focused on the intrinsic magnetic topological insulators because the interaction between magnetism and topological surface states can produce many exotic topological quantum effects, such as the quantum anomalous Hall effect (QAHE), Majorana bound states and axion insulator states \cite{8Mong,9ZhangDongqin,10ChangCui-Zu,11DengYujun,12XuYuanfeng,13QiXiaoLiang,14LiHang,15GuiXin,16Mogi,17GongYan,18ChenLi,19ZhouLiqin,20LiJiaheng,21Otrokov}. Recently, EuSn$_2$As$_2$ has been confirmed to be an intrinsic magnetic topological insulator by combining first-principles calculation and angle resolved photoemission spectroscopy (ARPES) measurements \cite{14LiHang}. EuSn$_2$As$_2$ crystallizes in a Bi$_2$Te$_3$-type structure in rhombohedral ($R\bar{3}m$) symmetry and consists of SnAs bilayers sandwiched by six coordinated Eu cations via van der Waals (vdW) bonding. The weak vdW force between the bilayers allows EuSn$_2$As$_2$ to be readily exfoliated into few layer sheets \cite{22Arguilla}.
Meanwhile, magnetic susceptibility measurements suggested that EuSn$_2$As$_2$ undergoes an antiferromagnetic (AFM) transition with Neel temperature \textit{T}${\rm {_N}}$ $\sim 24$ K, where the Eu$^{2+}$ ions are coupled ferromagnetically within each layer and antiferromagnetically across the adjacent SnAs bilayers, forming an A-type AFM order \cite{14LiHang,22Arguilla}.

\begin{figure*}
\includegraphics[width=14.0cm]{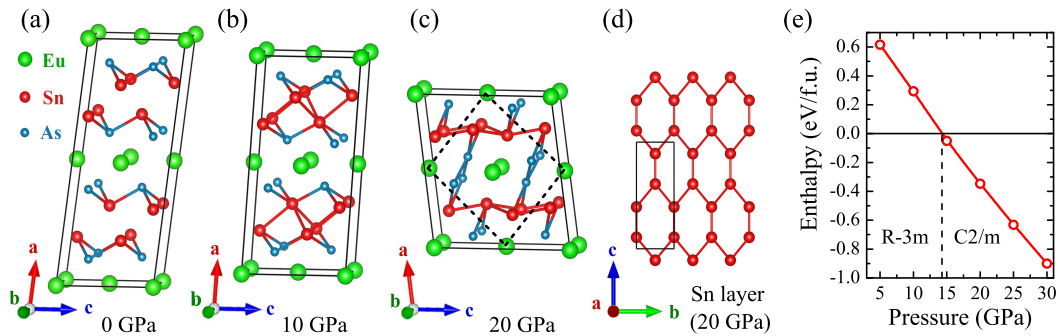}
\caption{
(a) $\alpha$-EuSn$_2$As$_2$ in rhombohedral $R\bar{3}m$ (No. 166) symmetry at 0 GPa with four buckled As-Sn layers and two Eu layers.
(b) $\alpha$-EuSn$_2$As$_2$ in rhombohedral $R\bar{3}m$ symmetry at 10 GPa with the connecting of Sn-Sn atoms between the nearest neighbor SnAs layers.
(c) $\beta$-EuSn$_2$As$_2$ in monoclinic $C2/m$ (No. 12) symmetry at 20 GPa with the connecting of As-As atoms between the SnAs layers across the Eu layers.
It has a network structure comprising honeycomb-like Sn sheets and zigzag As chains.
The magnetic unit cell in $P2/m$ (No. 10) symmetry is marked with dashed lines.
(d) Top view of the honeycomb-like Sn sheet in (c).
(e) Relative enthalpy between the $\alpha$- and $\beta$-EuSn$_2$As$_2$ phases as a function of pressure.
	}
\end{figure*}

Pressure, as a basic thermodynamic parameter, plays an important role in the researches of topological materials,
and can effectively tune the crystal and electronic structure of material to form a new state of matter.
The pressure-induced superconductivity has been successfully observed in the typical topological insulators Bi$_2$Se$_3$ \cite{24Kirshenbaum,25KongPp}, Bi$_2$Te$_3$ \cite{26,27ZhangC,28Matsubayashi,29ZhangJunliang} and Sb$_2$Te$_3$ \cite{30ZhuJ}, which crystallize in a rhombohedral structure at ambient pressure and undergoes structural phase transitions toward a monoclinic phase under high pressure \cite{prb184110}. The structural transitions mostly correspond to the appearance of different superconducting phases or \textit{T}${\rm {_C}}$ mutations, illustrating the inevitable correlation between structure and superconductivity \cite{25KongPp,26,31ZhangSJ,32Einaga,33ZhuLil,34ZhaoJinggeng,35Feng,36ZhangJ}.
Meanwhile, EuSn$_2$As$_2$ has a similar Bi$_2$Te$_3$-type layered structure in rhombohedral symmetry under ambient pressure, however, a systematic high-pressure study on EuSn$_2$As$_2$ is still lacking.

In this Letter, we present a comprehensive study on the structural phase transition and electrical transport property of EuSn$_2$As$_2$ under a wide pressure range of 0$-$30 GPa.
Combining {\it ab initio} calculations and \emph{in-situ} x-ray diffraction measurements, we identify a monoclinic EuSn$_2$As$_2$ phase in $C2/m$ symmetry above $\sim$14 GPa.
This new-type monoclinic phase has a network structure comprising honeycomb-like Sn layers and zigzag As chains and
can be transformed from the layered rhombohedral structure via a two-stage reconstruction mechanism with the connecting of Sn-Sn and As-As atoms successively between the buckled SnAs layers.
Its dynamic structural stability has been verified by phonon mode analysis.
Detailed electrical resistance measurements reveal an insulator-metal-superconductor transition at low temperature around 5 and 15 GPa, respectively, correspond to the structural conversion process.
These results show a strong correlation between the structure and electrical transport property under pressure

Our density functional theory (DFT) calculations are performed using the Vienna $ab$ $initio$ simulation package \cite{377vasp} with the projector augmented wave method \cite{388paw}.
The Perdew-Burke-Ernzerhof revised for solids (PBEsol) \cite{399ps} exchange-correlation functional is
adopted for the evaluation of structural and magnetic stability of EuSn$_2$As$_2$ under pressure.
The valence states 5s$^2$6s$^2$5p$^6$4f$^7$ for Eu, 5s$^2$5p$^2$ for Sn, and 4s$^2$4p$^3$ for As are used
for the plane wave basis set.
To match the energy position of Eu 4$f$ bands in the experiments \cite{14LiHang}, the Hubbard U = 5 eV is used to treat the localized 4$f$ electrons of Eu in the DFT+U scheme \cite{400plusU}.
Phonon calculations are performed using the PHONOPY package \cite{ph1}.
Meanwhile, our {\it in-situ} high-pressure experiments are performed in a diamond anvil cell (DAC) 
with culet size of 300 $\mu$m in diameter. 
The synchrotron x-ray diffraction (XRD) patterns are collected with wavelength of 0.6199 \AA\ and integrated using FIT2D software \cite{fit2d}.
The electrical resistance measurements are established by the standard four-probe methods at the Synergic Extreme Condition User Facility \cite{Yuxiaohui}.
Detailed computational and experimental methods are given in Supplemental Material \cite{433SM}.

\begin{figure}[b]
	\includegraphics[width=8.50cm]{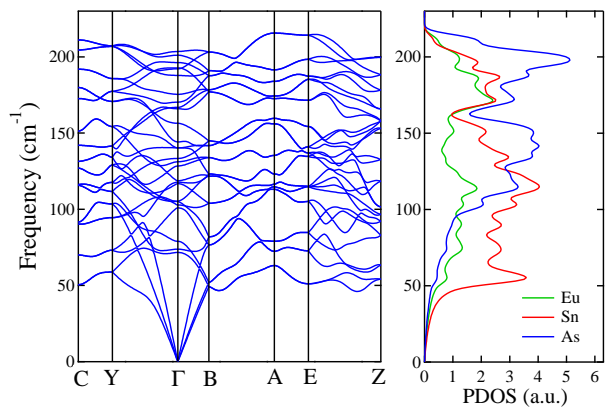}
	\caption{Phonon band structures and PDOS for $\beta$-EuSn$_2$As$_2$ at 20 GPa in a monoclinic ($P2/m$) magnetic primitive cell.
}
\end{figure}

We first characterize the structural phase conversion of the layered EuSn$_2$As$_2$ under pressure.
In order to understand the phase transition mechanism and determine the high-pressure structure,
we have examined various conversion processes with the reconstruction between SnAs layers as the finding in graphite \cite{W} by {\it ab initio} calculations.
We pay special attention to the initial geometry of rhombohedral structure (termed $\alpha$-EuSn$_2$As$_2$ hereafter) and adopt a simple monoclinic supercell (including 4 Eu, 8 Sn, and 8 As atoms), as shown in Fig. 1(a),
to simulate the interlayer AFM interaction of Eu atoms in the layered $\alpha$-phase \cite{14LiHang}.
Our calculations show that the more stable structure above 14.3 GPa is a monoclinic structure in $C2/m$ (No. 12) symmetry [see Fig. 1(c), termed $\beta$-EuSn$_2$As$_2$ hereafter].
The enthalpy of $\beta$-phase relative to the $\alpha$-phase is shown in Fig. 1(e) as a function of pressure up to 30 GPa.
One can easily catch sight of a two-stage phase transition process under pressure.
At the first-stage, from structure (a) [see Fig. 1(a)] toward structure (b) [see Fig. 1(b)],
the two buckled SnAs layers connect to each other via the nearest neighbor Sn-Sn atoms,  while the structures keep well in rhombohedral symmetry;
at the second-stage, from structure (b) toward structure (c) [see Fig. 1(c)],
the buckled Sn-Sn bonds become planar and form honeycomb-like Sn sheets [see Fig. 1(d)], meanwhile the SnAs layers further connect to each other via the As-As bonds across the Eu layers to form zigzag As chains between the Sn sheets.
As a result, a three-dimensional monoclinic network structure comprising honeycomb-like Sn sheets and zigzag As chains is achieved
via a two-stage reconstruction mechanism.

It is worth noting that the $\beta$-EuSn$_2$As$_2$ phase has a monoclinic magnetic primitive cell [marked by dashed line in Fig. 1(c)] in $P2/m$ (No. 10) symmetry  with alternating interlayer and intralayer AFM coupling between the Eu atoms.
The calculated magnetic moments on Eu sites are 6.92-6.98 $\mu_B$ due to the 4$f$ electrons.
The lattice parameters at 20 GPa in $P2/m$ symmetry are estimated to be 
$a$ = 7.659\AA,  $b$ = 3.596\AA,  $c$ = 6.958\AA, and $\beta$ = 87.33$^o$, occupying
1$a$ (0.0, 0.0, 0.0)-Eu1, 1$h$ (0.50, 0.50, 0.50)-Eu2,
2$m$ (0.3882, 0.0, 0.1704)-Sn1, 2$n$ (0.8882, 0.5, 0.6704)-Sn2,
2$m$ (0.7860, 0.0, 0.3730)-As1, and 2$n$ (0.2860, 0.5, 0.8730)-As2 Wyckoff positions.
In order to confirm the dynamical stability, we have calculated the phonon band structures and partial density of states (PDOS) in $P2/m$ symmetry at 20 GPa. As shown in Fig. 2,
the high frequency models around 200 and 140 cm$^{-1}$ are mainly contributed by As atoms,
while the low frequency modes around 55 cm$^{-1}$ are mainly contributed by Sn atoms.
Throughout the entire Brillouin zone, no imaginary frequencies are observed, confirming the dynamic stability of this new $\beta$-EuSn$_2$As$_2$ phase.

\begin{figure}[b]
	\includegraphics[width=8.50cm]{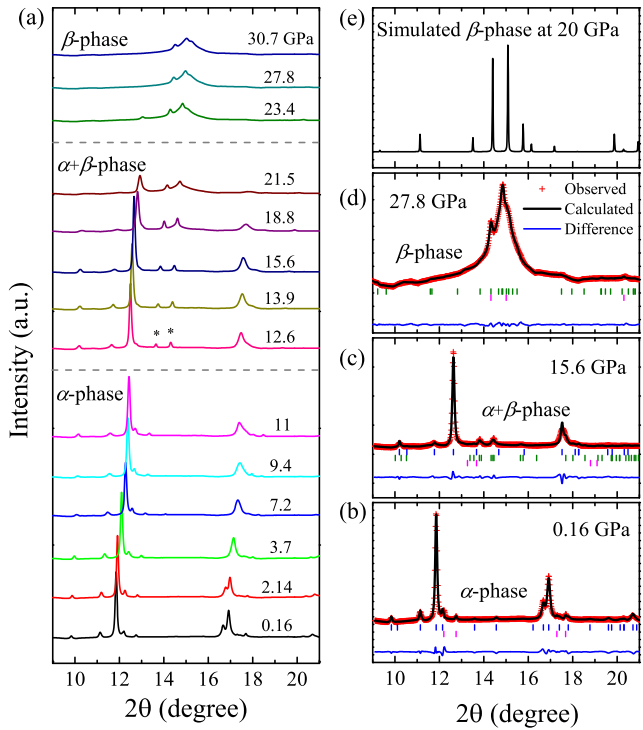}
	\caption{(a) The \textit{in-situ} XRD patterns of EuSn$_2$As$_2$ under 0.16$-$30.7 GPa.
		The synchrotron x-ray wavelength $\lambda$ = 0.6199 \AA.
		(b-d) The refined XRD patterns at 0.16 GPa, 15.6 GPa, and 27.8 GPa. The blue, green, and pink vertical tick markers are corresponding to the Bragg reflections of $\alpha$-phase, $\beta$-phase, and Sn, respectively.
		(e) Simulated XRD patterns of $\beta$-phase at 20 GPa. The two main peaks are in corresponding to
		the new diffraction peaks in (a) above 12.6 GPa.
	}
\end{figure}

\begin{table}
	\caption{Calculated lattice parameters at 0, 10, and 20 GPa for $\alpha$-EuSn$_2$As$_2$ in rhombohedral ($R\bar{3}m$) and $\beta$-EuSn$_2$As$_2$ in monoclinic ($C2/m$) symmetry,
		compared to our and the reported experimental data \cite{22Arguilla} at 0.16, 11, and 18.8 GPa.
	}
	\begin{tabular}{lllcccc} \hline
		\hline
		Structure  & Method           &  $a$ (\AA) & $b$ (\AA) &  $c$ (\AA) &  $\beta$ ($^o$)  &  $P$ (GPa)   \\
		\hline
		$\alpha$-EuSn$_2$As$_2$   & Cal    & 4.202 & 4.202 & 26.157 &   &  0  \\
		& Cal                              & 4.055 & 4.055 & 24.246   &  & 10    \\
		& Exp\cite{22Arguilla}  & 4.207 & 4.207  & 26.473 &   &  0  \\
		& Exp  & 4.213 & 4.213  & 26.354 &   &  0.16  \\
		
		& Exp  & 4.112  & 4.112 & 24.589   &   &  11   \\
		
		\hline
		$\beta$-EuSn$_2$As$_2$  & Cal & 10.584 & 3.596 & 7.659 & 138.95  & 20   \\
		& Exp   & 10.928  & 3.474  & 8.242  & 138.69  &  18.8   \\
		
		\hline
		\hline
	\end{tabular}
\end{table}

Figure 3(a) shows the experimental XRD patterns of EuSn$_2$As$_2$ up to 30.7 GPa at room temperature.
In the low-pressure range below 11 GPa, no obvious structure changes are observed in the XRD patterns,
indicating that the rhombohedral $\alpha$-phase persists up to 11 GPa.
With increasing pressure to 12.6 GPa, two new notable diffraction peaks at 14$\sim$15$^o$ appear and gradually higher with further compression, manifesting the occurrence of a structural phase transition.
Meanwhile, the main peaks of $\alpha$-phase around 12$^o$ and 17$^o$ gradually weakened and almost disappeared at 23.4 GPa,
indicating the phase transition almost completely achieves around 23.4 GPa.
In order to get the best understanding of the experimental data, we have refined the powder XRD pattern by Rietveld method \cite{Rietveld} through the Fullprof software \cite{fullprof2}.
At 0.16 GPa [see Fig. 3(b)], almost all of the diffraction peaks can be well indexed to the rhombohedral $\alpha$-phase
with space group $R\bar{3}m$, except for tiny diffraction peaks which are attributable to the residual Sn flux with a small weight fraction of $\sim$ 0.59 $\%$ (see Fig. S1 in Supplemental Material \cite{433SM}).
Based on our proposed monoclinic $\beta$-EuSn$_2$As$_2$ structure and the simulated XRD patterns (see Fig. 3(e) at 20 GPa and Fig. S2 in Supplemental Material \cite{433SM}), we further performed Rietveld refinements on the experimental data at 15.6 GPa and 27.8 GPa.
As shown in Fig. 3(c) and 3(d), the calculated XRD patterns are in good agreement with the experimental XRD patterns.
The refinement \textit{R} factors at 15.6 GPa are \textit{R}$_p$ = 2.44$\%$ and \textit{R}$_{wp}$ = 3.15$\%$, at 27.8 GPa are \textit{R}$_p$ = 0.66$\%$ and \textit{R}$_{wp}$ = 0.93$\%$, respectively.
The calculated and experimental lattice parameters for both $\alpha$- and $\beta$-EuSn$_2$As$_2$ are listed in Table I
(details on Wyckoff positions are given in Table S1 in Supplemental Material \cite{433SM}), compared to available experimental data \cite{22Arguilla}.

\begin{figure*}[t]
	\includegraphics[width=16.00cm]{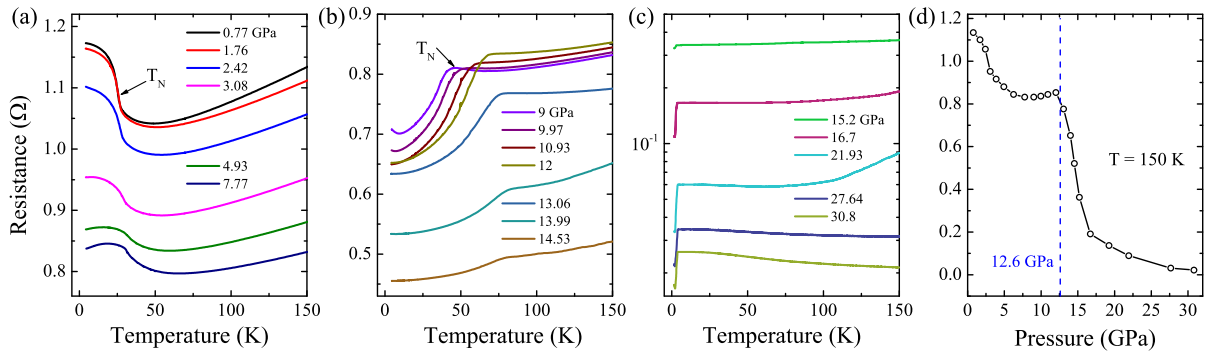}
	\caption{(a-c) Temperature dependent resistance of EuSn$_2$As$_2$ under various pressures up to 30.8 GPa.
		The insulator-metal-superconductor transition at low temperature are shown around 5 and 15 GPa, respectively.
The arrows indicate the Neel temperature \textit{T}${\rm {_N}}$.
(d) The resistivity-pressure curve at 150 K.
The two discontinuous change points near 5 and 12.6 GPa are in corresponding to the first-stage and second-stage structural conversion, respectively.
	}
\end{figure*}

We next discuss the electronic properties at low and high temperature range.
Figure 4(a-c) plots the resistance (\textit{R}) of EuSn$_2$As$_2$ single crystal as a function of temperature under various pressures up to 30.8 GPa.
For $0.77 \le P \le 3.08$ GPa, at low temperature below \textit{T}${\rm {_N}}$ [see Fig. 4(a)],
the resistance increases with the decrease of temperature, exhibiting an insulating behavior.
For  $4.93 \le P \le 7.77$, resistance increases first around \textit{T}${\rm {_N}}$, reaches a maximum then followed by a slight drop upon further cooling, exhibiting a metal-like behavior.
With increasing pressure above 9 GPa [see Fig. 4(b)],
the resistance decreases distinctly with decreasing temperature, exhibiting a typical metallic behavior below \textit{T}${\rm {_N}}$.
More interestingly, a superconducting transition with onset temperature \textit{T}${\rm {_C}}$ $\sim 4$ K is observed at 15.2 GPa [see Fig. 4(c)].
The superconductivity persists up to 30.8 GPa with \textit{T}${\rm {_C}}$ maintaining a constant value $\sim 4$ K.
To corroborate the superconductivity, we have also measured the temperature dependent resistance with various external magnetic fields (see Fig. S3 in Supplemental Material \cite{433SM}), and the upper critical fields (H$_{C2}$) is estimated of 0.89 T by using Ginzburg-Landau function.
On the other hand, at high temperature range, EuSn$_2$As$_2$ exhibits a metallic behavior, the resistivity-pressure curve (at 150 K) shows two discontinuous change points near 5 and 12.6 GPa [see Fig. 4(d)], which correspond to the first-stage and second-stage structural conversion, respectively.
To get a systematic understanding for the high-pressure behavior, we also map out a temperature-pressure phase diagram (see Fig. S4 in Supplemental Material \cite{433SM}).
We can see that the pressure has a remarkable influence on the structural and electrical behavior in EuSn$_2$As$_2$.

According to the electrical transport measurement,
one can also see that the antiferromagnetism associated with Eu$^{2+}$ moments changes during compression.
The \textit{T}${\rm {_N}}$ [marked by the arrows in Fig. 4(a,b)] can be determined by the minimum in the derivative of \textit{R}(T) \cite{45EuFeAs,46EuFeAs}.
As shown in Fig. 4(a) and 4(b), with increasing pressure, \textit{T}${\rm {_N}}$ monotonously increases from 25.4 K at 0.77 GPa to 86 K at 14 GPa.
Similar increasing behavior of \textit{T}${\rm {_N}}$ under pressure has been also observed in recently reported magnetic topological insulator EuIn$_2$As$_2$ \cite{45YuFH}, but no structural phase transitions and superconductivity are observed up to 17 GPa in the crystalline EuIn$_2$As$_2$.

In summary, we have identified a new monoclinic EuSn$_2$As$_2$ structure in $C2/m$ symmetry above $\sim$14 GPa
by {\it ab initio} calculations and \emph{in-situ} XRD measurements.
This distinct monoclinic phase has a three-dimensional network structure comprising honeycomb-like Sn sheets and zigzag As chains,
transformed from the layered rhombohedral structure via a two-stage reconstruction mechanism with the connecting of Sn-Sn and As-As atoms successively between the buckled SnAs bilayers.
Meanwhile, electrical transport measurements revealed an insulator-metal-superconductor transition at low temperature around 5 and 15 GPa, respectively, correspond to the two-stage structural conversion process.
These discoveries spread over the research fields of superconductivity, topological insulators, and quantum magnetism, and thus will provide a new venue for studying various topics of current condensed matter physics.

This work was supported by the National Key R$\&$D Program of China
(Grants No. 2016YFA0401503, No. 2016YFA0300604, No. 2018YFA0305700, No. 2018YFA0305900, and No. 2020YFA0711502),
the National Natural Science Foundation of China (Grant No. 11974387, No. 12004416, No. 11674328, No. U2032204, No. 11774126, No. 12004014, No. U1832123, and No. U1930401),
the Strategic Priority Research Program and Key Research Program of Frontier Sciences of the Chinese Academy of Sciences (Grant No. XDB33000000, No. XDB25000000, and No. QYZDBSSW-SLH013), the K. C. Wong Education Foundation (No. GJTD-2018-01), and the Youth Innovation Promotion Association of Chinese Academy of Sciences (No. 2016006).
The \emph{in-situ} XRD measurements were performed at 4W2 High Pressure Station, Beijing Synchrotron Radiation Facility (BSRF), which is supported by Chinese Academy of Sciences (Grant No.  KJCX2-SWN20, and No. KJCX2-SW-N03).


\begin{thebibliography}{99}

\bibitem{1Hasan}
M. Z. Hasan and C. L. Kane,
Rev. Mod. Phys. {\bf 82,} 3045 (2010).

\bibitem{2QiXiaoLiang}
X. L. Qi and S. C. Zhang,
Rev. Mod. Phys. {\bf 83,} 1057 (2011).


\bibitem{3FuLiang}
L. Fu and C. L. Kane,
Phys. Rev. Lett. {\bf 100,} 096407 (2008).

\bibitem{4Moore}
J. E. Moore,
Nature {\bf 464,} 194 (2010).

\bibitem{5ZhangHaijun}
H. J. Zhang, C. X. Liu, X. L. Qi, X. Dai, Z. Fang, and S. C. Zhang,
Nat. Phys. {\bf 5,} 438 (2009).

\bibitem{6FuLiang}
L. Fu and C. L. Kane,
Phys. Rev. B {\bf 76}, 045302 (2007).

\bibitem{7Kane}
C. L. Kane and E. J. Mele,
Phys. Rev. Lett. {\bf 95}, 146802 (2005).

\bibitem{8Mong}
R. S. K. Mong, A. M. Essin and J. E. Moore,
Phys. Rev. B {\bf 81}, 245209 (2010).

\bibitem{9ZhangDongqin}
D. Q. Zhang, M. J. Shi, T. S. Zhu, D. Y. Xing, H. J. Zhang, and J. Wang,
Phys. Rev. Lett. {\bf 122}, 206401 (2019).

\bibitem{10ChangCui-Zu}
C. Z. Chang, {\it et. al.,}
Science {\bf 340}, 167 (2013).

\bibitem{11DengYujun}
Y. J. Deng, Y. J. Yu, M. Z. Shi, Z. X. Guo, Z. H. Xu, J. Wang, X. H. Chen, and Y. B. Zhang,
Science {\bf 367}, 6480 (2020).

\bibitem{12XuYuanfeng}
Y. F. Xu, Z. D. Song, Z. J. Wang, H. Weng, and X. Dai,
Phys. Rev. Lett. {\bf 122}, 256402 (2019).

\bibitem{13QiXiaoLiang}
X. L. Qi, T. L. Hughes, and S. C. Zhang,
Phys. Rev. B {\bf 82,} 184516 (2010).

\bibitem{14LiHang}
H. Li, S. Y. Gao, S. F. Duan, Y. F. Xu, K. J. Zhu, S. J. Tian, J. C. Gao, W. H. Fan, Z. C. Rao,
J. R. Huang, J. J. Li, D. Y. Yan, Z. T. Liu, W. L. Liu, Y. B. Huang, Y. L. Li, Y. Liu, G. B. Zhang,
P. Zhang, T. Kondo, S. Shin, H. C. Lei, Y. G. Shi, W. T. Zhang, H. M. Weng, T. Qian, and H. Ding,
Phys. Rev. X {\bf 9}, 041039 (2019).

\bibitem{15GuiXin}
X. Gui,
I. Pletikosic, H. B. Cao, H. J. Tien, X. T. Xu, R. D. Zhong, G. Q. Wang, T. R. Chang, S. Jia, T. Valla, W. W. Xie, and R. J. Cava,
Acs. Central. Sci. {\bf 5}, 900 (2019).

\bibitem{16Mogi}
M. Mogi, M. Kawamura, R. Yoshimi, A. Tsukazaki, Y. Kozuka, N. Shirakawa, K. S. Takahashi, M. Kawasaki, and Y. Tokura,
Nat. Mater. {\bf 16}, 516 (2017).

\bibitem{17GongYan}
Y. Gong, {\it et. al.,}
Chinese Phys. Lett. {\bf 36}, 089901 (2019).

\bibitem{18ChenLi}
L. Chen, D. C. Wang, C. M. Shi, C. Jiang, H. M. Liu, G. L. Cui, X. M. Zhang, and X. L. Li,
J. Mater. Sci. {\bf 55}, 14292 (2020).

\bibitem{19ZhouLiqin}
L. Q. Zhou, Z. Y. Tan, D. Y. Yan, Z. Fang, Y. G. Shi, and H. Weng,
Phys. Rev. B {\bf 102}, 085114 (2020).

\bibitem{20LiJiaheng}
J. Li, C. Wang, Z. Zhang, B. L. Gu, W. Duan, and Y. Xu,
Phys. Rev. B {\bf 100}, 121103(R) (2019).

\bibitem{21Otrokov}
M. M. Otrokov, I. P. Rusinov, M. Blanco-Rey, M. Hoffmann, A. Y. Vyazovskaya, S. V. Eremeev, A. Ernst, P. M. Echenique, A. Arnau, and E. V. Chulkov,
Phys. Rev. Lett. {\bf 122}, 107202 (2019).

\bibitem{22Arguilla}
M. Q. Arguilla, N. D. Cultrara, Z. J. Baum, S. Jiang, R. D. Ross, and J. E. Goldberger,
Inorg. Chem. Front. {\bf 4}, 378 (2017).

\bibitem{24Kirshenbaum}
K. Kirshenbaum, P. S. Syers, A. P. Hope, N. P. Butch, J. R. Jeffries, S. T. Weir, J. J. Hamlin, M. B. Maple, Y. K. Vohra, and J. Paglione,
Phys. Rev. Lett. {\bf 111}, 087001 (2013).

\bibitem{25KongPp}
P. P. Kong, J. L. Zhang, S. J. Zhang, J. Zhu, Q. Q. Liu, R. C. Yu, Z. Fang, C. Q. Jin, W. G. Yang, X. H. Yu, J. L. Zhu, and Y. S. Zhao,
J. Phys.: Condens. Matter {\bf 25}, 362204 (2013).

\bibitem{26}
J. L. Zhang, {\it et. al.,}
Proc. Natl. Acad. Sci. U.S.A. {\bf 108}, 24 (2011).

\bibitem{27ZhangC}
C. Zhang, L. Sun, Z. Chen, X. Zhou, Q. Wu, W. Yi, J. Guo, X. Dong, and Z. Zhao,
Phys. Rev. B {\bf 83}, 140504(R) (2011).

\bibitem{28Matsubayashi}
K. Matsubayashi, T. Terai, J. S. Zhou, and Y. Uwatoko,
Phys. Rev. B {\bf 90}, 125126 (2014).

\bibitem{29ZhangJunliang}
J. L. Zhang, S. J. Zhang, P. P. Kong, L. X. Yang, C. Q. Jin, Q. Q. Liu, X. C. Wang, and J. C. Yu,
J. Appl. Phys. {\bf 123}, 125901 (2018).

\bibitem{30ZhuJ}
J. Zhu, {\it et. al.,}
Sci. Rep. {\bf 3}, 2016 (2013).

\bibitem{prb184110}
R. Vilaplana, D. Santamaria-Perez, O. Gomis, F. J. Manjon, J. Gonzalez, A. Segura, A. Munoz, P. Rodriguez-Hernandez,
E. Perez-Gonzalez, V. Marin-Borras, V. Munoz-Sanjose, C. Drasar, and V. Kucek,
Phys. Rev. B {\bf 84,} 184110 (2011).

\bibitem{31ZhangSJ}
S. J. Zhang, {\it et. al.,}
J. Appl. Phys. {\bf 111}, 1757 (2012).

\bibitem{32Einaga}
M. Einaga, F. Ishikawa, A. Ohmura, A. Nakayama, Y. Yamada, and S. Nakano,
Phys. Rev. B {\bf 83}, 092102 (2011).

\bibitem{33ZhuLil}
L. Zhu, H. Wang, Y. C. Wang, J. A. Lv, Y. M. Ma, Q. L. Cui, Y. M. Ma, and G. T. Zou,
Phys. Rev. Lett. {\bf 106}, 145501 (2011).

\bibitem{34ZhaoJinggeng}
J. G. Zhao, H. Z. Liu, L. Ehm, Z. Q. Chen, S. Sinogeikin, Y. S. Zhao, and G. D. Gu,
Inorg. Chem. {\bf 50}, 11291 (2011).

\bibitem{35Feng}
F. Ke, {\it et. al.,}
Adv. Mater. {\bf 29}, 1701983 (2017).

\bibitem{36ZhangJ}
J. L. Zhang, S. J. Zhang, J. L. Zhu, Q. Q. Liu, X. C. Wang, C. Q. Jin, and J. C. Yu,
Physica B {\bf 521}, 13 (2017).

\bibitem{377vasp}
G. Kresse and J. Furthm$\ddot{u}$ller,
Phys. Rev. B {\bf 54}, 11169 (1996).


\bibitem{388paw}
P. E. Bl\"{o}chl,
Phys. Rev. B {\bf 50}, 17953 (1994).



\bibitem{399ps}
J. P. Perdew, A. Ruzsinszky, G. I. Csonka, O. A. Vydrov, G. E. Scuseria, L. A. Constantin, X. Zhou, and K. Burke,
Phys. Rev. Lett. {\bf 100,} 136406 (2008).


\bibitem{400plusU}
V. I. Anisimov, J. Zaanen, and O. K. Andersen,
Phys. Rev. B {\bf 44,} 943 (1991).


\bibitem{ph1}
A. Togo, F. Oba, and I. Tanaka,
Phys. Rev. B {\bf 78}, 134106 (2008).

\bibitem{fit2d}
A. P. Hammersley, S. O. Svensson, M. Hanfland, A. N. Fitch, and D. Hausermann,
High Pressure Res. {\bf 14}, 235 (1996).

\bibitem{Yuxiaohui}
X. H. Yu, F. F. Li, Y. H. Han, F. Hong, C. Q. Jin, Z. He, and Q. Zhou,
Chinese Phys. B {\bf 27}, 070701 (2018).


\bibitem{433SM}
See Supplemental Material for the detailed computational and experimental methods;
the single crystal and powdered XRD pattern of EuSn$_2$As$_2$ at ambient pressure (Fig. S1);
the simulated XRD pattern for $\alpha$- and $\beta$-EuSn$_2$As$_2$ at various pressure (Fig. S2);
the refined lattice parameters and atomic coordinates for EuSn$_2$As$_2$ at various pressure (Table S1);
the temperature-dependent resistance at various magnetic filed (Fig. S3);
and the temperature-pressure phase diagram of EuSn$_2$As$_2$ (Fig. S4).

\bibitem{W}
J. T. Wang, C. F. Chen, and Y. Kawazoe,
Phys. Rev. Lett. {\bf 106}, 075501 (2011).

\bibitem{Rietveld}
H. M. Rietveld,
J. Appl. Cryst. {\bf 2}, 65 (1969).

\bibitem{fullprof2}
J. Rodriguezcarvajal,
Physica B {\bf 192}, 55 (1993).

\bibitem{45EuFeAs}
K. Matsubayashi, K. Munakata, M. Isobe, N. Katayama, K. Ohgushi, Y. Ueda, Y. Uwatoko, N. Kawamura, M. Mizumaki, N. Ishimatsu, M. Hedo, and I. Umehara,
Phys. Rev. B {\bf 84}, 024502 (2011).

\bibitem{46EuFeAs}
N. Kurita, M. Kimata, K. Kodama, A. Harada, M. Tomita, H.S. Suzuki, T. Matsumoto, K. Murata, S. Uji, and T. Terashima,
Phys. Rev. B {\bf 83}, 214513 (2011).


\bibitem{45YuFH}
F. H. Yu, H. M. Mu, W. Z. Zhuo, Z. Y. Wang, Z. F. Wang, J. J. Ying, and X. H. Chen,
Phys. Rev. B {\bf 102}, 180404(R) (2020).


\end{thebibliography}
\end{document}